\documentclass[amsmath,amssymb,amsfonts,aps,prb,twocolumn,superscriptaddress]{revtex4-2}

\usepackage{color}
\usepackage{graphicx}
\usepackage[capitalize]{cleveref}
\usepackage{braket}
\usepackage[normalem]{ulem}

\begin{document}
	
\newcommand{\eqdef}{\overset{\gamma_{iL}=\gamma_{iR}=\frac{\gamma_i}{2}}{=\joinrel=\joinrel=\joinrel=\joinrel=\joinrel=\joinrel=}}
\newcommand{\w}{{\omega}}
\newcommand{\g}{{\gamma}}
	
\title{Level occupation switching with density functional theory}

\author{Nahual Sobrino} \email{nahualcsc@dipc.org}
\affiliation{Donostia International Physics Center (DIPC), Paseo Manuel de
  Lardizabal 4, E-20018 San Sebasti\'{a}n, Spain}
\affiliation{Departamento de Pol\'{i}meros y Materiales Avanzados: F\'{i}sica, Qu\'{i}mica y Tecnolog\'{i}a, Universidad del Pa\'{i}s Vasco UPV/EHU,
  Av. Tolosa 72, E-20018 San Sebasti\'{a}n, Spain}

\author{David Jacob}\email{david.jacob@ehu.es}
\affiliation{Departamento de Pol\'{i}meros y Materiales Avanzados: F\'{i}sica, Qu\'{i}mica y Tecnolog\'{i}a, Universidad del Pa\'{i}s Vasco UPV/EHU,
  Av. Tolosa 72, E-20018 San Sebasti\'{a}n, Spain}
\affiliation{IKERBASQUE, Basque Foundation for Science, Plaza Euskadi 5, E-48009 Bilbao, Spain}
 
\author{Stefan Kurth}
\affiliation{Departamento de Pol\'{i}meros y Materiales Avanzados: F\'{i}sica, Qu\'{i}mica y Tecnolog\'{i}a, Universidad del Pa\'{i}s Vasco UPV/EHU,
  Av. Tolosa 72, E-20018 San Sebasti\'{a}n, Spain}
\affiliation{IKERBASQUE, Basque Foundation for Science, Plaza Euskadi 5, E-48009 Bilbao, Spain}
\affiliation{Donostia International Physics Center (DIPC), Paseo Manuel de
  Lardizabal 4, E-20018 San Sebasti\'{a}n, Spain}
 
\begin{abstract}
  The charge transport properties of zero-temperature multi-orbital quantum
  dot systems with one dot coupled to leads and the other dots coupled only
  capacitatively are studied within density functional theory. It is shown that
  the setup is equivalent to an effective single impurity Anderson model. This allows to understand the level occupation switching effect as transitions
  between ground states of different integer occupations in the uncoupled dots.
  Level occupation switching is very sensitive to small energy differences and
  therefore also to the details of the parametrized exchange-correlation
  functionals. An existing functional already captures the effect on a
  qualitative level but we also provide an improved parametrization which is 
  very accurate when compared to reference numerical renormalization group
  results. 
\end{abstract}

 \date{\today}
 \maketitle
 
\section{Introduction}

Quantum dots (QDs) are an ideal testbed to investigate the interplay between
quantum many-body physics and transport phenomena. They can be fabricated
in the lab from a large variety of materials and techniques, such as metallic nanoparticles\cite{Petta2001},
lateral confinement of a two-dimensional electron gas (2DEG)
in GaAs/AlGaAs heterostructures (for a review see Ref.~\onlinecite{Hanson2007} and references therein),
carbon nanotubes (CNTs)\cite{Nygaard2000,Buitelaar2002}, and
molecular junctions\cite{Park2002}.
Indeed important many-body phenomena such as the Kondo effect\cite{Hewson_Book}
and Coulomb blockade (CB)\cite{Averin1986}, characteristic for so-called strongly correlated electrons,
where electronic interactions dominate over the kinetic energy, have been
measured in transport setups of QDs.\cite{Shtrikman1998,Goldhaber1998,Park2002,Champagne2005,Parks2007}

The physics of QDs can be further enriched by the existence of multiple electronic levels (or orbitals),
or by coupling of QDs. The interplay between strong electronic correlations and the spin and orbital degrees of freedom in
multi-orbital QDs, may lead to new physical phenomena, such as the SU(4) and underscreened Kondo effects, which have
both been measured in CNTs.\cite{Jarillo2005,Roch2009,Parks2010}
An interesting effect may occur in multi-orbital QDs when one of the QD levels couples more strongly to the leads than
the other levels. In this case abrupt changes in the conductance and transmission phases between Coulomb blockade peaks
have been observed.\cite{avinun2005crossover,yacoby1995coherence,schuster1997phase,aikawa2004interference,ji2002transmission,zaffalon2008transmission} These may be attributed to the so-called level occupation switching (LOS), where
the strongly coupled level is abruptly emptied, while the weakly coupled level(s) are abruptly filled, or vice versa. \cite{lim2006kondo,kleeorin2017abrupt,busser2011transport,roura2011interplay,weymann20184,silvestrov2007level}
In essence this phenomenon is a result of the competition between the kinetic energy of the strongly coupled level and its
electrostatic repulsion with the weakly coupled levels.\cite{busser2011transport}

Density functional theory (DFT) is one of the most successful and popular approaches for computing the electronic structure
of molecules and solids owing to its relative simplicity and computational efficiency.\cite{HohenbergKohn:64,Kohn:PR:1965,DreizlerGross:90}
Since DFT is in principle an exact (many-body) theory for the ground-state energy and density of a many-electron system, it should
also be capable of describing strong electronic correlation phenomena such as Kondo effect, Coulomb blockade, and ultimately the LOS effect.
However, in practice approximations need to be made in DFT for the exchange-correlation (xc) part of the total energy functional.
And unfortunately, the most popular approximations to DFT, such as the local-density\cite{Kohn:PR:1965} and generalized-gradient
approximations\cite{Perdew:85,becke1988_gga,perdew1996generalized} in condensed-matter physics, and the so-called hybrid functionals
in chemistry,\cite{Becke:93-2} are known to fail for strongly correlated systems.


  
Nevertheless, if equipped with proper approximations for the xc part of the
functional, DFT is indeed capable of describing strongly correlated
phenomena such as Coulomb blockade and Kondo effect in transport through
nanoscale devices,\cite{sk.2011,blbs.2012,tse.2012}
and the Mott-Hubbard gap in solids.\cite{LimaOliveiraCapelle:02} More
recently, it has also been shown that the actual many-body spectral function
may be extracted from a DFT calculation by making use of an extension of DFT
called i-DFT.\cite{JacobKurth:18,kurth2019nonequilibrium} This DFT
framework also allows to describe the Mott metal-insulator transition, one of
the hallmarks of strong electronic correlations.\cite{JacobStefanucciKurth:20}
The crucial ingredient for the description of these phenomena within DFT are
steps at integer occupations in the xc potentials.\cite{sobrino2020exchange}
These steps are related to the famous derivative discontinuity of exact DFT
\cite{PerdewParrLevyBalduz:82}, which is missing in the standard
approximations. In the context of the Anderson impurity model, the step
feature in the xc potential gives rise to the pinning of the Kohn-Sham (KS)
impurity level to the Fermi energy, which results in a plateau for the
zero-bias conductance as a function of the applied gate, in accordance with
the Kondo effect.\cite{DreizlerGross:90,Perdew:85,becke1988_gga}
  
Here we show how DFT can be used to study the LOS phenomenon which occurs in multi-level quantum dot ({\rm MQD}) systems in the asymmetric
situation where only one of the levels is connected to leads. In
\cref{sec_model} we introduce the model and show how it can exactly be mapped 
onto an effective single impurity Anderson model (SIAM).
In \cref{sec_lattice_dft} we use lattice DFT for the effective SIAM and an
energy minimization argument to decide which configuration of integer
occupations of the disconnected dots is realized. In \cref{sec_results} we
show that an existing parametrization of the SIAM Hxc potential already
qualitatively captures the LOS effect although not always at the correct
value of the gate. The origin of these deviations is investigated and remedied
by a re-parametrization of the Hxc functional. Finally, we present our
conclusions in \cref{sec_conclus}.



\section{Model}
\label{sec_model}  

\begin{figure}
  \includegraphics[width=0.9\linewidth]{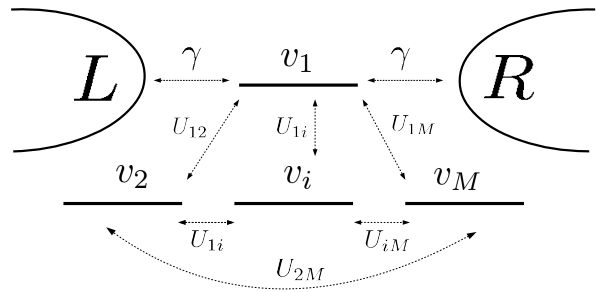}
  \caption[DQD transport setup]{Schematic representation of the transport setup for the {\rm MQD}. The first impurity is connected to the electron reservoirs while the rest only interact through electrostatic repulsions with each other.}
  \label{fig_setup_MQD}
\end{figure}

We consider a multi-orbital quantum dot consisting of
$M$ impurities which can hold up to two electrons and which are all 
capacitively coupled among each other. 
We consider the situation when only one of the impurities is
connected to two leads and we also restrict ourselves to the zero-temperature
limit. The total Hamiltonian of the system is given as the sum of the
Hamiltonians of the isolated dot and leads, as well as the coupling between
them and reads

\begin{align}
\label{eq_H}
\hat{H} = \hat{H}_{{\rm MQD}} + \sum_{\alpha k \sigma} \varepsilon_{\alpha k}
\hat{c}^{\dagger}_{\alpha k \sigma} \hat{c}_{\alpha k \sigma} \nonumber\\
+ \sum_{\alpha k \sigma} \left( t_{\alpha k} \hat{d}^{\dagger}_{1 \sigma}
\hat{c}_{\alpha k \sigma} + {\rm H.c.} \right)
\end{align}
where 
\begin{align}
\label{eq_H_MQD}
\hat{H}_{{\rm MQD}} = \sum_{i=1}^{M} v_{i}\hat{n}_{i}+
  \sum_{i=1}^{M}U_{i}\hat{n}_{i\uparrow}\hat{n}_{i\downarrow} +
    \sum_{i<j}^{M} U_{ij}\hat{n}_{i}\hat{n}_{j}
\end{align}
describes the capacitatively coupled {\rm MQD} where
$\hat{n}_i=\hat{n}_{i\uparrow}+\hat{n}_{i\downarrow}$ is the number operator for
level $i$ with $\hat{n}_{i\sigma}=\hat{d}_{i\sigma}^{\dagger}\hat{d}_{i\sigma}$ and
$\hat{d}_{i\sigma}$ ($\hat{d}_{i\sigma}^\dagger$) are the annihilation (creation)
operators for electrons
with spin $\sigma$ in orbital $i$. In \cref{eq_H_MQD}, $v_{i}$ and $U_{i}$
are the on-site energy (also referred to as gate) and the intra-Coulomb
repulsion of level $i$ while $U_{ij}$ is the inter-Coulomb repulsion between
levels $i$ and $j$. The second term in \cref{eq_H} describes the
non-interacting electrons in left (L) and right (R) leads ($\alpha=L,R$)
while the third term accounts for the (symmetric) coupling of the first
impurity to left and right leads. The resulting broadening functions
$\Gamma_{\alpha}(\w) = 2 \pi \sum_k |t_{\alpha k}|^2 \delta(\w - \varepsilon_{\alpha k})$
are assumed to describe featureless leads and therefore become independent of
frequency, i.e., we take the wide band limit (WBL) with
$\Gamma_{\alpha}(\w) = \gamma_{\alpha}$. In \cref{fig_setup_MQD} a 
schematic representation of the {\rm MQD} setup is shown.

Since the impurities  $j=2,\ldots,M$ are not connected to the
reservoirs, the multi-orbital quantum dot system can be mapped exactly onto
an effective SIAM.\cite{busser2011transport} This follows from the fact
that the operators $\hat{n}_{j \sigma}$ for $j>1$ all commute with the
Hamiltonian of \cref{eq_H}. Therefore, all the many-body eigenstates of
$\hat{H}$ can be chosen to be eigenstates of $\hat{n}_{j \sigma}$ (for $j>1$)
and the corresponding eigenvalues take values $n_{j \sigma}=0,1$. Therefore,
the total occupations $n_j$ of the disconnected dots can only take the integer
values $n_j=0,1,2$. As a consequence of this commutation property, when
the Hamiltonian is applied to an eigenstate of $\hat{n}_{j \sigma}$ ($j>1$), the
problem is seen to be equivalent to an effective SIAM related to the
first impurity with an effective potential 
\begin{align}
\label{eq_eff_pot}
v_{1}^{\rm eff} = v_{1}+\sum_{j=2}^{M} U_{1j} n_j 
\end{align}
and an additional constant contribution to the total energy given by
$\sum_{j\ne 1}U_{j}\delta_{n_j,2}$. Therefore, the only effect of charging and
discharging the levels $j=2,\ldots,M$ is to modify the average
Hartree potential felt by the first level. 

\section{Lattice density functional theory}
\label{sec_lattice_dft}

With the observations of the previous Section it is clear that any many-body
method which can accurately treat the SIAM can be employed to obtain the
ground state energy and density of $\hat{H}$. Here our method of choice is
lattice DFT which in the past has successfully been used for the SIAM.
\cite{sk.2011,blbs.2012,tse.2012,KurthStefanucci:16}.

\subsection{Energy functional}

In lattice DFT, for a given set of gates
$\mathbf{v}= (v_1,\ldots,v_M)$ the total energy functional for the
interacting system described by $\hat{H}$ reads
\begin{align}
\label{eq_Energy}
E[\mathbf{n}] = T_s[n_1] + E_{\rm Hxc}[\mathbf{n}] + \sum_{i=1}^{M} n_iv_i,
\end{align}
where $T_{s}[n_1]$ is the non-interacting kinetic energy of the only impurity
connected to leads (the kinetic energy of the disconnected dots vanishes).
$E_{\rm Hxc}[\mathbf{n}]$ is the Hartree-exchange-correlation (Hxc) energy
which is a functional of all densities and we used the notation
$\mathbf{n}= (n_1,\ldots,n_M)$. The non-interacting kinetic energy
can be expressed in terms of the (local) Green function $G^R_{d}(\omega)$ of
the connected dot as $T_{s}=\frac{\gamma}{\pi}\int_{-\varepsilon_C}^{0}d\omega
\text{Re}\left[G_{d}^R(\omega)\right]$.
Here $\varepsilon_C$ is an energy cutoff which ensures the convergence of the
integral. In the WBL the non-interacting kinetic energy contribution can be
expressed in a closed form (see Appendix for the detailed derivation) as
\begin{align}
  \label{KS_kinetic}
T_{s}[n]= \frac{\gamma}{2\pi}\log\left[\frac{\frac{\gamma^{2}}{4}\left(\tan{\left(\frac{\pi}{2}(1-n)\right)}^{2}+1\right)}{\varepsilon_C^2+\frac{\gamma^2}{4}}\right].
\end{align}
Note that the energy cutoff only acts as a constant shift to the total energy
but does not change its shape. Therefore both the energy minimum and any
energy difference between different configurations are independent of the
value of $\varepsilon_C$.

The Hxc energy functional $E_{\rm Hxc}[\mathbf{n}]$ of the total system can be
simplified using the considerations of the previous Section. Since for the
disconnected dots $i\in\{2 ,\ldots ,M\}$, the only possible
occupations are $n_i =0,1,2$ (and the KS system has to reproduce these
occupations), $E_{\rm Hxc}$ may be written as 
\begin{align}
\label{Hxc_energy}
E_{\rm Hxc}[\mathbf{n}] = E_{\rm Hxc}^{\rm SIAM}[n_1] + \sum_{i<j}^{M}
U_{ij} n_i n_j + \sum_{j=2}^{M} U_{j} \delta_{n_j,2}
\end{align}
where $E_{\rm Hxc}^{\rm SIAM}[n]$ is the Hxc functional for a simple SIAM.
Inserting \cref{Hxc_energy} into \cref{eq_Energy} it becomes immediately
clear that the resulting total energy functional has the form of the energy
of a single Anderson impurity but with effective on-site potential
$v_1^{\rm eff}$ given by \cref{eq_eff_pot}. 

For a given set of gate levels ($v_1,\ldots,v_M$), there are
$3^{M-1}$ different configurations of occupations
($n_2,\ldots,n_M$) of the disconnected dots. The ground state energy
and the resulting set of occupancies of the system can then be obtained by
comparing the energies of the different configurations of available states and
taking the one corresponding to the minimum of energy.
In the degenerate case $v_i=v$, the ground state energy of the system can be found by minimizing the universal functional $F[n]=T_{s}[n] + E_{Hxc}[n]$, since the last term of \cref{eq_Energy} is constant (at given fixed total occupation) due to the one-to-one correspondence between total occupation $N$ and the external potential $v$.

A LOS event exactly corresponds to a change in the ground state energy of the
system, and provided an accurate parametrization for the SIAM Hxc energy
$E_{\rm Hxc}^{\rm SIAM}$ is used, this event can be completely captured within DFT.

\subsection{Kohn-Sham equation}

For a given configuration of (integer) occupations ($n_2,\ldots,n_M$)
of the disconnected dots, the (non-integer) density on the connected dot can
now be found by the Hohenberg-Kohn variational principle, i.e., by searching
for the value $n_1$ which minimizes the total energy (\ref{eq_Energy}).
Therefore we need to solve $\frac{\partial E[\mathbf{n}]}{\partial n_1}=0$
which is easily shown to be equivalent to solving the KS equation
\begin{align}
\label{eq_density_KS}
n_1 =1-\frac{2}{\pi}\arctan{\left(2\frac{v_{1}^{\rm eff} + v_{\rm Hxc}^{\rm SIAM}[n_1]}
  {\gamma_1}\right)},
\end{align}
where we have used the definition (\ref{eq_eff_pot}) of the effective
potential $v_1^{\rm eff}$ and defined the SIAM Hxc potential as
\begin{align}
\label{SIAM_Hxc_pot}
v_{\rm Hxc}^{\rm SIAM}[n] = \frac{{\rm d} E_{\rm Hxc}^{\rm SIAM}[n]}{{\rm d} n} \;. 
\end{align}

We note in passing that \cref{eq_density_KS} is equivalent to expressing
the density as\cite{kurth2017transport}
\begin{align}
\label{eq_density}
n_{1} =2\int_{-\infty}^0 \frac{{\rm d}\w}{2 \pi} \;A_{s,1}(\w)
\end{align}
where, without loss
of generality, we assumed a vanishing equilibrium chemical potential in the
leads. The KS spectral function for the connected dot is
\begin{align}
\label{_eq_A_KS}
A_{s,1}(\omega)=\frac{\g_{1}}{\frac{\g_{1}^{2}}{4}+(\omega-v_{s,1})^{2}} \;,
\end{align}
consistent with WBL approximation used to derive \cref{KS_kinetic}, and $v_{s,1} = v_{1}^{\rm eff} + v_{\rm Hxc}^{\rm SIAM}$.

  According to Friedel sum rule, in the zero-temperature limit, the impurity 
spectral function at the Fermi energy is completely determined by the impurity
density.\footnote{See, e.g., Ch.~5.2 in the book by Hewson\cite{Hewson_Book}}
Since in the setup considered here only one impurity level is connected to the leads,
the zero-bias conductance is directly given by the spectral function at the Fermi level.
In this case the Friedel sum rule implies that the electrical conductance of the system is
fully determined by the equilibrium density at the impurity and reads\cite{mera2010assessing,langreth1966friedel,mera2010kohn}
\begin{align}
\label{eq_FSR}
G = \sin{\left(\frac{\pi n_{1}}{2}\right)}^{2}.
\end{align} 
On the other hand, the correct description of the electrical conductance requires the access to the actual (many-body) electrical current
of the system, and therefore one possibility is the inclusion of the xc
corrections to the bias of the system.
\cite{StefanucciKurth:15,sobrino2021thermoelectric,sobrino2019steady}
However, since here the exact zero-bias conductance can be expressed explicitly
in terms of the ground state density alone (which is a quantity accessible to
standard ground state DFT), already the KS zero-bias conductance $G_s$ is exact
\cite{sk.2011,blbs.2012,tse.2012,kurth2016nonequilibrium,stefanucci2013kondo}
and can be expressed as
\begin{align}
G = G_s =  \frac{\frac{\gamma^{2}}{4}}{v_{s,1}^{2}+\frac{\gamma^{2}}{4}}.
\label{eq_G_low_T}
\end{align}
Eqs.(\ref{eq_FSR}) and (\ref{eq_G_low_T}) provide two equivalent expressions
in the zero temperature regime for the zero-bias conductance. \cite{mera2010assessing,langreth1966friedel,mera2010kohn}

\begin{figure}
	\includegraphics[width=0.9\linewidth]{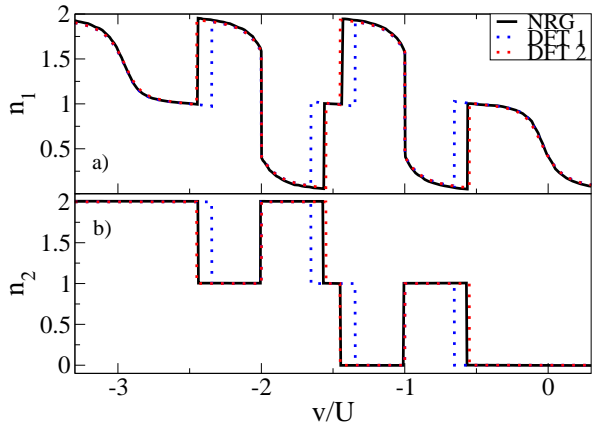}
	\caption[]{Comparison of the local occupancies obtained with  the
          exchange-correlation functional of
          Ref.~\onlinecite{blbs.2012} (DFT 1), and those obtained with the new parametrization of \cref{sigma_new} (DFT 2) to the NRG
          results of Ref.~\onlinecite{kleeorin2017abrupt}. Both $n_i$ are
          shown as function of the gate level for the degenerate case
          $v=v_1=v_2$ and for strong correlations $\gamma/U=0.1$.}
	\label{fig_ni}
\end{figure}

\section{Results}
\label{sec_results}

As discussed in Section \ref{sec_model}, the problem of the capacitively
coupled {\rm MQD} with only one impurity coupled to leads can exactly
be mapped onto a SIAM. In the context of lattice DFT
(Sec.~\ref{sec_lattice_dft}) this implies that the only quantity to be
approximated is the Hxc functional $E_{\rm Hxc}^{\rm SIAM}$ from which
$v_{\rm Hxc}^{\rm SIAM}$ follows by differentiation. Once such a parametrization
is given, the {\rm MQD} DFT problem is solved as follows: For a given set
of gate potentials $\mathbf{v}$, for each configuration of (integer)
occupations of the disconnected impurities (covering the $3^{M-1}$
possible configurations) we solve the KS equation (\ref{eq_density_KS}) for
$n_1$ with the effective potential of \cref{eq_eff_pot} and calculate
the corresponding total energies. The lowest of these energies then
corresponds to the configuration of the ground state.


For the SIAM functional, we start by considering the new parametrization at zero temperature suggested in Ref.~\onlinecite{blbs.2012}.
In \cref{fig_ni} we present the local occupancies for the case of the double
quantum dot $M=2$ as a function of the gate level in the degenerate
case $v=v_1=v_2$ and strong correlations $\gamma/U=0.1$. The results labeled
DFT 1 correspond to the self-consistent densities obtained with the
parametrization of Ref.~\onlinecite{blbs.2012}. The first thing to
note is that our approach does capture the LOS transitions and gives
densities which qualitatively agree with the reference NRG results
of Ref.~\onlinecite{kleeorin2017abrupt}. However, we also notice that the LOS
events take place at values of $v$ considerably different from the
many-body results. Since, as mentioned above, the only possible source of
error in our approach is the parametrization of $E_{\rm Hxc}^{\rm SIAM}$, below we
propose a reparametrization of the functional (DFT 2 in \cref{fig_ni}) which
correctly and accurately captures the LOS transitions.

\begin{figure}
  \includegraphics[width=1\linewidth]{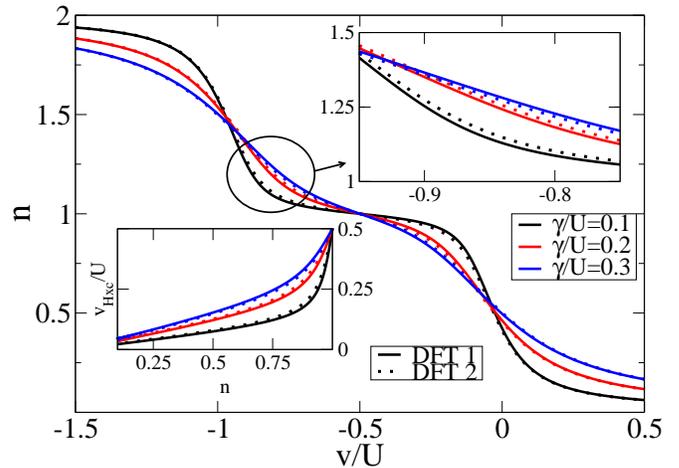} 
  \caption{Self-consistent densities of the simple SIAM obtained with
      the parametrization of $v_{\rm Hxc}^{\rm SIAM}$ given in
      Ref.~\onlinecite{blbs.2012} (DFT 1) and our new parametrization
      (DFT 2) as function of gate for different values of $\gamma/U$. The
      upper insets highlights small differences in the mixed-valence regime
      while the lower inset shows the Hxc potentials for the two
      parametrizations.}
  \label{fig_functionals}
\end{figure}

\begin{figure}
	\includegraphics[width=1\linewidth]{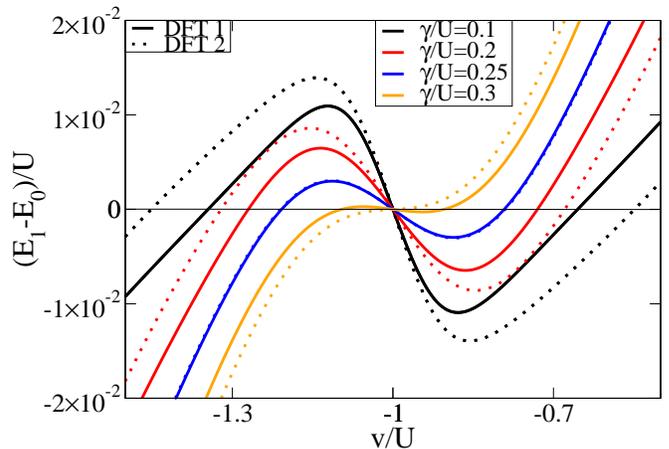} 
	\caption{Energy difference between the states with $n_2=0$ and $n_2=1$ of
		the double quantum dot for different coupling strengths in the case
		$v=v_1=v_2$. The LOS events exactly correspond to the degeneracy of the
		states $E_{n_2=1}=E_{n_2=0}$. Note how the new parametrization shifts the crossings of zero
		of the energy differences, i.e., the gate values at which the LOS event occurs.}
	\label{fig_energy_cross}
\end{figure}

The parametrization of Ref.~\onlinecite{blbs.2012} depends on
two parameters which are both functions of $\gamma/U$. In order to correctly
capture the LOS events we here propose to keep the same functional form but
reparametrize the parameter $\sigma$ of Eq.~(16) of
Ref.~\onlinecite{blbs.2012}. Our fit to those numerical values of
$\sigma$ which best reproduces the positions of the LOS events for the
parameters of Ref.~\onlinecite{kleeorin2017abrupt} is given as
\begin{align}
\label{sigma_new}
\sigma = 0.07\arctan\left(\frac{171.358 (\gamma/U)^2} 
       {2+\gamma/U}\right) \;.
\end{align}
This newly parametrized Hxc functional for the SIAM will be denoted as DFT 2
in the following. It now accurately captures the gate positions of the LOS events, see Fig.~\ref{fig_ni}. 


The gate positions of the LOS events are highly sensitive to the
parametrization of the Hxc functional, especially in the mixed valence regime.
This can be appreciated in
\cref{fig_functionals}, where we show the self-consistent
SIAM densities produced with the DFT 1 (solid line) and the DFT 2 (dashed
line) functionals for different coupling strengths. The discrepancies between
the corresponding densities are almost negligible, with the maximum
disagreement in the mixed valence regime, i.e., in the transition from empty
to half occupation and from half occupation to full occupation. In the inset
of \cref{fig_functionals}, both parametrizations of
$v_{\rm Hxc}^{\rm SIAM}[n]$ are compared. Note that although the corrections are
very small (of the order of $\sim 10^{-3}U$), they are crucial in order to
accurately capture the LOS events.

\begin{figure}
	\includegraphics[width=1\linewidth]{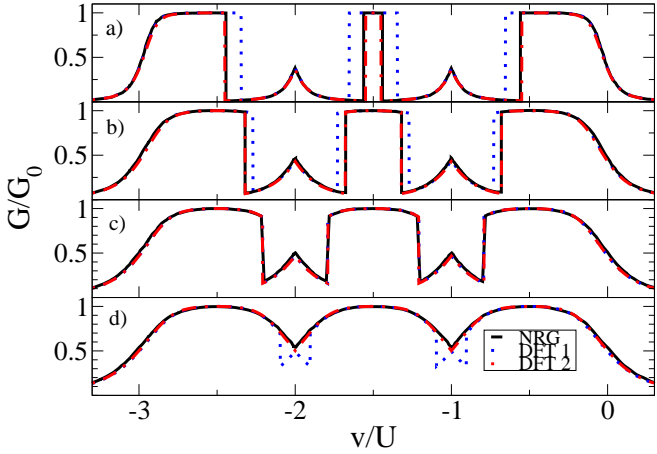}
	\caption[]{Comparison of the zero-bias conductance obtained with the
		exchange-correlation functional of Ref.~\onlinecite{blbs.2012}
		(DFT 1) with the new parametrization (DFT 2) and NRG calculations. $G$ is
		shown as function of the gate level in the degenerate case $v=v_1=v_2$ and
		for different coupling strengths $\gamma/U=0.1,0.2,0.25,0.3$ from a) to
		d), respectively. All energies in units of U.}
	\label{fig_G_dif_gamma}
\end{figure}

 In \cref{fig_energy_cross}, we show as an illustrative example
the difference between the computed energies for two different (integer)
$n_2$, $E_{n_2=1}$ and $E_{n_2=0}$, obtained self-consistently with the different
parametrizations. The effect of the new parametrization is considerably larger
for some values of the coupling strength ($\gamma/U=0.1,0.2,0.3$), while for
others it is not appreciable ($\gamma/U=0.25$). In particular, this difference
is relevant at $E_{n_2=1}-E_{n_2=0}=0$, which exactly corresponds to the gate at
which the the LOS event occurs. Although not shown in the present paper, the
new parametrization does not introduce any change in the $E_{n_2=2}=E_{n_2=0}$
crossing (only found at $v/U=-1,-2$) while the transition from $E_{n_2=2}$
to $E_{n_2=1}$ follows exactly the same correction as the one illustrated
for gates centered at $v/U=-2$. 

\subsection{Results for the double quantum dot}

\begin{figure}
	\includegraphics[width=1\linewidth]{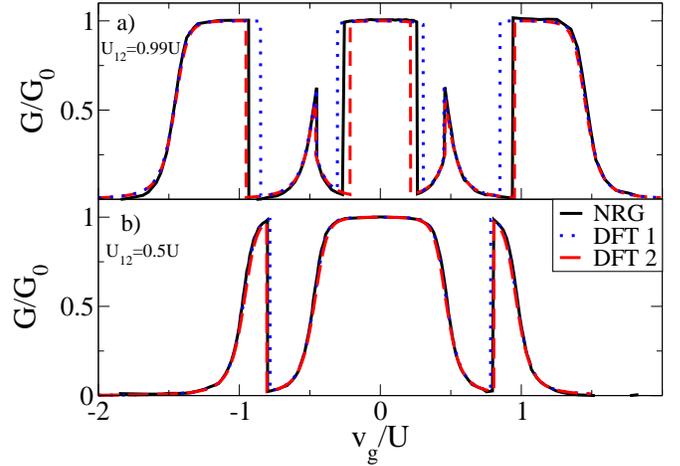}
	\caption[]{Zero-bias conductance as function of the gate voltage $v_g=v-U_{12}-U/2$ in units
		of $U$ for a)  $U_{12}=0.99U$ and b) $U_{12}=0.5U$. In both panels
		$\gamma/U=0.1$.}
	\label{fig_DQD_dif_U}
\end{figure}

In \cref{fig_G_dif_gamma} we present the conductance as function of the gate
level in the degenerate case $v=v_1=v_2$ for different coupling strengths
$\gamma/U=0.1,0.2,0.25,0.3$ from a) to d), respectively. For small coupling
strength, six LOS events occur, two of them pinned at the gate values
$v/U=-2,-1$, related to two peaks usually referred to as CB peaks,\cite{Kleeorin:PRB:2017} since similar structures are present in the CB regime. 
The other four LOS transitions lead to a widening of the three Kondo plateaus. We observe that the DFT 1 results are qualitatively correct, while still
failing to predict the evolution of the LOS events with increasing coupling
strength, except for the ones related to the two CB peaks.
On the other hand, the DFT 2 results accurately reproduce the NRG calculations,
showing the correct evolution of the three Kondo plateaus to wider structures
and the two CB peaks into dips for higher values of $\gamma/U=0.3$.

\begin{figure}
  \includegraphics[width=1\linewidth]{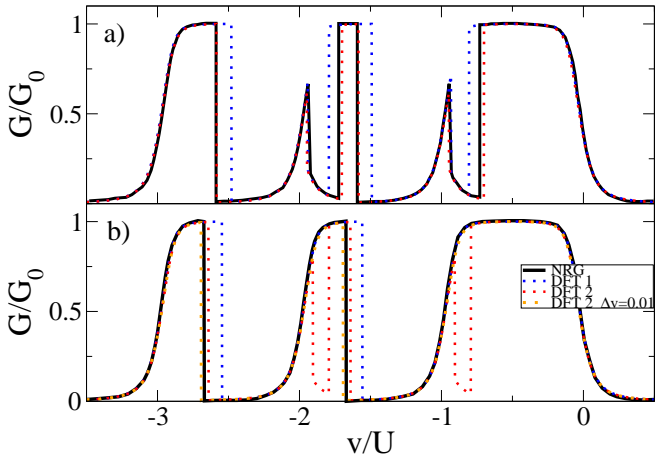}
  \caption[]{Conductance as function of the gate level in units of $U$ for a)  $\Delta v=0.005$ and b) $\Delta v=0.0075$. In both panels $\gamma/U=0.1$.}
  \label{fig_DQD_dif_v}
\end{figure}
 
The results shown so far are related to the fully degenerate case, i.e,
$v=v_1=v_2$ and $U=U_1=U_2=U_{12}$, but our DFT study can go further. Since the
dependence on the inter-Coulomb repulsions only enters into the effective
potential felt by the first impurity, we can easily investigate the effects
of changing it. In \cref{fig_DQD_dif_U}  we show the conductances for
a) $U_{12}=0.99U$ and b) $U_{12}=0.5U$ as function of the gate voltage
$v_g=v-U_{12}-U/2$ in units of $U$. Again, we find that DFT 2 gives an
excellent agreement with NRG except for the central plateau in panel a),
where the NRG calculations predict a slightly wider central structure. When
the inter-Coulomb repulsion is decreased to be half of the intra-Coulomb
repulsion, the two CB peaks  and the central Kondo
plateau merge into a smooth plateau leading to the disappearance of four
of the LOS events. On the other hand, in \cref{fig_DQD_dif_v} we explore the
effect of considering a finite difference between the gate levels
$\Delta v=v_1-v_2$. In \cref{fig_DQD_dif_v} a) the DFT 2 parametrization
correctly captures the evolution of the two CB peaks 
for $\Delta v=0.005$, while for $\Delta v=0.0075$ (\cref{fig_DQD_dif_v} b))
the DFT 2 results present a finite difference with the reference NRG results.
The origin of this discrepancy is not completely clear: it could either be
due to subtle details of the Hxc potentials which our parametrization
does not capture. However, it could also be due to  small numerical effects due to
the sensitivity of the LOS events to small energy differences.
 However, we do find that
our results for $\Delta v=0.01$ completely agree with the NRG ones for $\Delta v=0.0075$.

\subsection{Results for more than two dots}

We further apply our new parametrization of $v_{\rm Hxc}^{\rm SIAM}$ to the
situation of more than two orbitals with only one of them being connected to leads.
The generalization is straightforward and only requires to find the ground
state energy between the different $3^{M-1}$ states corresponding
to the different configurations of integer occupancies of the disconnected
dots.

\begin{figure}
	\includegraphics[width=1\linewidth]{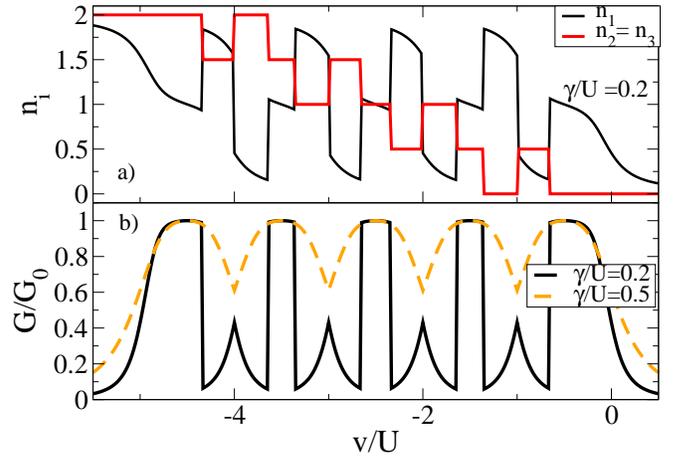}
	\caption[]{Densities and electrical conductance of the triple quantum dot as function of the gate level for the fully degenerate case and $\gamma/U=0.2$.}
	\label{fig_triple_qd}
\end{figure}

Some results for the fully degenerate case of the triple quantum dot
($v=v_1=v_2=v_3$) for $\gamma/U=0.2$ are presented in \cref{fig_triple_qd}.
In panel a) the densities predict a total of twelve LOS events. Since both
the second and third impurity levels are completely degenerate, a swap in
their local occupancies leaves the problem invariant and the corresponding
many-body eigenstates are degenerate. Therefore, the (average) occupancies
may take semi-integer values for some values of the gate. Following the same
reasoning we observe that for the fully degenerate {\rm MQD},
the $M-1$ degenerate levels of disconnected dots can only reach
occupancies of integer multiples of $(M-1)^{-1}$. In panel b) the
conductance present four CB peaks and five Kondo plateaus,
which evolve with the coupling strength in an analogous manner but reaching
the inversion of the CB peaks into valleys
around $\gamma/U\sim0.5$.

\begin{figure}
	\includegraphics[width=1\linewidth]{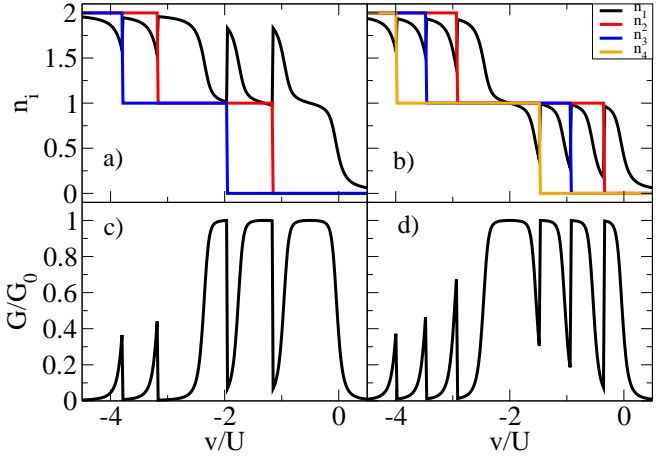}
	\caption[]{DFT  densities and zero-bias conductance as function of the gate level. Panels a) and c): Triple quantum dot with $U=\frac{5}{4}U_{12}=\frac{5}{3}U_{13}=2U_{23}$   and $v=v_1=v_2-0.1=v_3-0.4$. Panels b) and d):  Quadruple quantum dot with $U=U_i=2U_{ij}$ for all $i\ne j$ and $v=v_1=v_2-0.02=v_3-0.02=v_4-0.04$. In both cases we chose $U=10\gamma$.}
	\label{fig_triple_quad_results}
\end{figure}

Finally, in \cref{fig_triple_quad_results} we present the densities and the
related conductances for the triple (panels a) and c)) and quadruple
(panels b) and d)) quantum dots in the nondegenerate case. For the triple
quantum dot we choose the values $U=\frac{5}{4}U_{12}=\frac{5}{3}U_{13}=2U_{23}$   and
$v=v_1=v_2-0.1=v_3-0.4$  and for the quadruple quantum dot $U=U_i=2U_{ij}$ for
all $i\ne j$ and $v=v_1=v_2-0.02=v_3-0.02=v_4-0.04$. In both cases we choose
$U=10\gamma$. For the selected set of parameters, the LOS events always correspond to an abrupt filling of a weakly coupled impurity and an abrupt emptying of the strongly coupled one. In both systems we observe that as the gate is decreased, a LOS event for $n_1<1$ ($n_1>1$) implies a sudden decreasing (increasing) of the conductance.

 \section{Conclusions}
 \label{sec_conclus}

 In this work we have studied from a DFT perspective the LOS effect which
 occurs in multi-orbital quantum dots subject to inter and intra Coulomb
 repulsions when only one of the dots is coupled to leads. The system can be
 mapped into an effective SIAM problem for the coupled impurity which
 experiences an effective gate potential due to electrostatic interactions
 with the other impurities.
The density of the system is obtained by choosing the minimum total energy
(expressed in terms of DFT quantities) among the available $3^{M-1}$
different configurations of integer occupation of the uncontacted levels. A
LOS event occurs at that gate for which the energy minimum passes from one
configuration of integer occupations to another one.
We have modified an already quite accurate parametrization of the
SIAM Hxc functional in order to correctly capture the coupling strength
dependence of the LOS events which is quite sensitive to details of the functional. Our new parametrization yields very small energy
differences (of the order of $\sim 10^{-3}U$) as compared to the previous one. 
This produces almost no effect on the selfconsistent densities of SIAM, but is
essential to shift the gates at which the LOS events occur to the correct
positions.  DFT calculations employing the parametrized Hxc functionals for
the double quantum dot show excellent agreement with many-body NRG
calculations. We have further presented results for the triple and quadruple quantum
dot.

\section{Acknowledgments}

We acknowledge funding through the  grant “Grupos Consolidados UPV/EHU del
Gobierno Vasco” (IT1453-22). We also acknowledges funding through a grant of
the ”Ministerio de Ciencia y Innovaci\'on (MCIN)” (Grant No. PID2020-112811GB-I00).


\appendix
\label{Appendix_1}
\section{Kinetic energy of a non-interacting impurity coupled to two leads}
The purpose of this Appendix is to derive the explicit functional of the
non-interacting kinetic energy of a non-interacting impurity connected to
left (L) and right (R) leads. We start by writing the Hamiltonian of this 
non-interacting impurity in second quantized form as
\begin{align}
\label{eq_H_appendix} 
\hat{H}=v_{s}\hat{d}^{\dagger}\hat{d} +\sum_{\alpha=L,R}
  \sum_{k\sigma}\epsilon_{\alpha k}\hat{c}^{\dagger}_{\alpha k\sigma}
  \hat{c}_{\alpha k\sigma}+\hat{H}_{\rm kin}
\end{align}
where
\begin{align}
  \hat{H}_{\rm kin} =\sum_{\alpha=L,R} \sum_{ k\sigma}\left( V_{\alpha k,d}
  \hat{d}^{\dagger}_{\sigma}\hat{c}_{\alpha k\sigma}+{\rm H.c.} \right)\; .
\end{align}

The (non-interacting) kinetic energy is then given as
\begin{align}
  \label{Ts_kin}
  T_s = \langle \hat{H}_{\rm kin} \rangle =
  2 \sum_{\alpha k} \left( - i V_{\alpha k,d} G_{\alpha k,d}^{<}(t^+,t) + {\rm c.c.}
  \right) 
\end{align}
where the prefactor two comes from spin
\begin{align}
  G_{\alpha k,d}^{<}(t,t') = i \langle \hat{d}^{\dagger}_{\sigma, H}(t')
  \hat{c}_{\alpha k \sigma, H}(t) \rangle
\end{align}
is the (spin-independent) matrix element of the lesser Green function between
single-particle basis states $|\alpha k \sigma \rangle$ and $|d \sigma\rangle$.
By standard Green function techniques \cite{stefanucci2013nonequilibrium} 
\cref{Ts_kin} can be written as 
\begin{align}
 \lefteqn{T_s = 
  2 \sum_{\alpha k} \left( V_{\alpha k,d} \int_{-\varepsilon_C}^0 \frac{{\rm d} \omega}{2 \pi}
  A_{\alpha k,d}(\omega) + {\rm c.c.} \right) }\nonumber \\
  &= 2 \sum_{\alpha k} \left( V_{\alpha k,d} \int_{-\varepsilon_C}^0
  \frac{{\rm d} \omega}{2 \pi} i (G_{\alpha k,d}^{R}(\omega) -
  G_{\alpha k,d}^{A}(\omega)) + {\rm c.c.} \right) \nonumber\\
  \label{Ts_kin2}
\end{align}
where $G^R$ ($G^A$) are the retarded (advanced) Green functions and we
introduced an energy cutoff $\varepsilon_C$ in order for the integral to
converge. The retarded Green function $\hat{G}^{R}(\omega)$ at energy $\omega$
is defined through
\cite{stefanucci2013nonequilibrium,kurth2017transport}
\begin{align}
\label{eq_G_def}
\left((\omega+i\eta)\hat{\mathbb{I}}-\hat{H}\right)\hat{G}^R(\omega)
=\hat{\mathbb{I}}.
\end{align}
In the single-particle basis, all Hamiltonian matrix elements directly
connecting left and right leads vanish, i.e.,
$\langle L k \sigma |\hat{H}| R k' \sigma' \rangle = \langle R k \sigma
|\hat{H}| L k' \sigma'\rangle = 0$.
Then, from \cref{eq_G_def} one can derive 
\begin{gather}
\label{eq_G_kd}
G_{\alpha k,d}^R(\omega)  = \frac{V_{\alpha k,d}^*}{\omega-\epsilon_k + i \eta }
G_{d}(\omega)\\
\label{eq_G_d}
G_{d}^R(\omega) = \frac{1}{\omega-v_s-\Delta^R(\omega)}
\end{gather}
with $\eta \to 0^+$ and $\Delta^R(\omega)= \sum_{\alpha}
\Delta_{\alpha}^R(\omega)$ is the total embedding self energy with
\begin{align}
  \label{delta_a}
  \Delta_{\alpha}^R(\omega) = \sum_{k} \frac{|V_{\alpha k, d}|^2}{\omega -
    \varepsilon_{\alpha k} + i \eta}
\end{align}
In the wide-band limit we have $\Delta_{\alpha}^R(\omega) = - i
\frac{\gamma_{\alpha}}{2}$, independent of $\omega$. Inserting
Eqs.~(\ref{eq_G_kd}) and (\ref{delta_a}) into \cref{Ts_kin2} we arrive at
\begin{align}
  \label{Ts_kin3}
  T_s = \frac{\gamma}{\pi} \int_{-\varepsilon_C}^0 {\rm d} \omega \;
  {\rm Re}[G_d^R(\omega)]
  = \frac{\gamma}{2 \pi} \log\left[\frac{v_s^{2}+\frac{\gamma^{2}}{4}}{\varepsilon_C^2+\frac{\gamma^2}{4}}\right]
\end{align}
The non-interacting density-potential relation is
\begin{align}
  n = 1 - \frac{2}{\pi} \arctan\left(\frac{2 v_s}{\gamma}\right)
\end{align}
which can easily be inverted to give
\begin{align}
  \label{vs_n_nonint}
  v_s = \frac{\gamma}{2} \tan\left(\frac{\pi}{2}(1-n)\right) \;. 
\end{align}
Inserting \cref{vs_n_nonint} into \cref{Ts_kin3} then gives the final result
for the non-interacting kinetic energy functional

\begin{align}
\label{eq_T_s_appendix}
T_s[n]=\frac{\gamma}{2\pi}
\log\left[\frac{\frac{\gamma^{2}}{4}\left(\tan{\left(\frac{\pi}{2}(1-n)\right)}^{2}+1\right)}{\varepsilon_C^2+\frac{\gamma^2}{4}}\right].
\end{align}



\begin{thebibliography}{55}%
	\makeatletter
	\providecommand \@ifxundefined [1]{%
		\@ifx{#1\undefined}
	}%
	\providecommand \@ifnum [1]{%
		\ifnum #1\expandafter \@firstoftwo
		\else \expandafter \@secondoftwo
		\fi
	}%
	\providecommand \@ifx [1]{%
		\ifx #1\expandafter \@firstoftwo
		\else \expandafter \@secondoftwo
		\fi
	}%
	\providecommand \natexlab [1]{#1}%
	\providecommand \enquote  [1]{``#1''}%
	\providecommand \bibnamefont  [1]{#1}%
	\providecommand \bibfnamefont [1]{#1}%
	\providecommand \citenamefont [1]{#1}%
	\providecommand \href@noop [0]{\@secondoftwo}%
	\providecommand \href [0]{\begingroup \@sanitize@url \@href}%
	\providecommand \@href[1]{\@@startlink{#1}\@@href}%
	\providecommand \@@href[1]{\endgroup#1\@@endlink}%
	\providecommand \@sanitize@url [0]{\catcode `\\12\catcode `\$12\catcode
		`\&12\catcode `\#12\catcode `\^12\catcode `\_12\catcode `\%12\relax}%
	\providecommand \@@startlink[1]{}%
	\providecommand \@@endlink[0]{}%
	\providecommand \url  [0]{\begingroup\@sanitize@url \@url }%
	\providecommand \@url [1]{\endgroup\@href {#1}{\urlprefix }}%
	\providecommand \urlprefix  [0]{URL }%
	\providecommand \Eprint [0]{\href }%
	\providecommand \doibase [0]{http://dx.doi.org/}%
	\providecommand \selectlanguage [0]{\@gobble}%
	\providecommand \bibinfo  [0]{\@secondoftwo}%
	\providecommand \bibfield  [0]{\@secondoftwo}%
	\providecommand \translation [1]{[#1]}%
	\providecommand \BibitemOpen [0]{}%
	\providecommand \bibitemStop [0]{}%
	\providecommand \bibitemNoStop [0]{.\EOS\space}%
	\providecommand \EOS [0]{\spacefactor3000\relax}%
	\providecommand \BibitemShut  [1]{\csname bibitem#1\endcsname}%
	\let\auto@bib@innerbib\@empty
	\bibitem [{\citenamefont {Petta}\ and\ \citenamefont
		{Ralph}(2001)}]{Petta2001}%
	\BibitemOpen
	\bibfield  {author} {\bibinfo {author} {\bibfnamefont {J.~R.}\ \bibnamefont
			{Petta}}\ and\ \bibinfo {author} {\bibfnamefont {D.~C.}\ \bibnamefont
			{Ralph}},\ }\href {\doibase 10.1103/PhysRevLett.87.266801} {\bibfield
		{journal} {\bibinfo  {journal} {Phys. Rev. Lett.}\ }\textbf {\bibinfo
			{volume} {87}},\ \bibinfo {pages} {266801} (\bibinfo {year}
		{2001})}\BibitemShut {NoStop}%
	\bibitem [{\citenamefont {Hanson}\ \emph {et~al.}(2007)\citenamefont {Hanson},
		\citenamefont {Kouwenhoven}, \citenamefont {Petta}, \citenamefont {Tarucha},\
		and\ \citenamefont {Vandersypen}}]{Hanson2007}%
	\BibitemOpen
	\bibfield  {author} {\bibinfo {author} {\bibfnamefont {R.}~\bibnamefont
			{Hanson}}, \bibinfo {author} {\bibfnamefont {L.~P.}\ \bibnamefont
			{Kouwenhoven}}, \bibinfo {author} {\bibfnamefont {J.~R.}\ \bibnamefont
			{Petta}}, \bibinfo {author} {\bibfnamefont {S.}~\bibnamefont {Tarucha}}, \
		and\ \bibinfo {author} {\bibfnamefont {L.~M.~K.}\ \bibnamefont
			{Vandersypen}},\ }\href {\doibase 10.1103/RevModPhys.79.1217} {\bibfield
		{journal} {\bibinfo  {journal} {Rev. Mod. Phys.}\ }\textbf {\bibinfo {volume}
			{79}},\ \bibinfo {pages} {1217} (\bibinfo {year} {2007})}\BibitemShut
	{NoStop}%
	\bibitem [{\citenamefont {Nyg{\aa}rd}\ \emph {et~al.}(2000)\citenamefont
		{Nyg{\aa}rd}, \citenamefont {Cobden},\ and\ \citenamefont
		{Lindelof}}]{Nygaard2000}%
	\BibitemOpen
	\bibfield  {author} {\bibinfo {author} {\bibfnamefont {J.}~\bibnamefont
			{Nyg{\aa}rd}}, \bibinfo {author} {\bibfnamefont {D.~H.}\ \bibnamefont
			{Cobden}}, \ and\ \bibinfo {author} {\bibfnamefont {P.~E.}\ \bibnamefont
			{Lindelof}},\ }\href {https://link.aps.org/doi/10.1103/PhysRevLett.88.156801}
	{\bibfield  {journal} {\bibinfo  {journal} {Nature}\ }\textbf {\bibinfo
			{volume} {408}},\ \bibinfo {pages} {342} (\bibinfo {year}
		{2000})}\BibitemShut {NoStop}%
	\bibitem [{\citenamefont {Buitelaar}\ \emph {et~al.}(2002)\citenamefont
		{Buitelaar}, \citenamefont {Bachtold}, \citenamefont {Nussbaumer},
		\citenamefont {Iqbal},\ and\ \citenamefont
		{Sch\"onenberger}}]{Buitelaar2002}%
	\BibitemOpen
	\bibfield  {author} {\bibinfo {author} {\bibfnamefont {M.~R.}\ \bibnamefont
			{Buitelaar}}, \bibinfo {author} {\bibfnamefont {A.}~\bibnamefont {Bachtold}},
		\bibinfo {author} {\bibfnamefont {T.}~\bibnamefont {Nussbaumer}}, \bibinfo
		{author} {\bibfnamefont {M.}~\bibnamefont {Iqbal}}, \ and\ \bibinfo {author}
		{\bibfnamefont {C.}~\bibnamefont {Sch\"onenberger}},\ }\href {\doibase
		10.1103/PhysRevLett.88.156801} {\bibfield  {journal} {\bibinfo  {journal}
			{Phys. Rev. Lett.}\ }\textbf {\bibinfo {volume} {88}},\ \bibinfo {pages}
		{156801} (\bibinfo {year} {2002})}\BibitemShut {NoStop}%
	\bibitem [{\citenamefont {Park}\ \emph {et~al.}(2002)\citenamefont {Park},
		\citenamefont {Pasupathy}, \citenamefont {Goldsmith}, \citenamefont {Chang},
		\citenamefont {Yaish}, \citenamefont {Petta}, \citenamefont {Rinkoski},
		\citenamefont {Sethna}, \citenamefont {Abru{\~{n}}a}, \citenamefont
		{McEuen},\ and\ \citenamefont {Ralph}}]{Park2002}%
	\BibitemOpen
	\bibfield  {author} {\bibinfo {author} {\bibfnamefont {J.}~\bibnamefont
			{Park}}, \bibinfo {author} {\bibfnamefont {A.~N.}\ \bibnamefont {Pasupathy}},
		\bibinfo {author} {\bibfnamefont {J.~I.}\ \bibnamefont {Goldsmith}}, \bibinfo
		{author} {\bibfnamefont {C.}~\bibnamefont {Chang}}, \bibinfo {author}
		{\bibfnamefont {Y.}~\bibnamefont {Yaish}}, \bibinfo {author} {\bibfnamefont
			{J.~R.}\ \bibnamefont {Petta}}, \bibinfo {author} {\bibfnamefont
			{M.}~\bibnamefont {Rinkoski}}, \bibinfo {author} {\bibfnamefont {J.~P.}\
			\bibnamefont {Sethna}}, \bibinfo {author} {\bibfnamefont {H.~D.}\
			\bibnamefont {Abru{\~{n}}a}}, \bibinfo {author} {\bibfnamefont {P.~L.}\
			\bibnamefont {McEuen}}, \ and\ \bibinfo {author} {\bibfnamefont {D.~C.}\
			\bibnamefont {Ralph}},\ }\href {\doibase 10.1038/nature00791} {\bibfield
		{journal} {\bibinfo  {journal} {Nature}\ }\textbf {\bibinfo {volume} {417}},\
		\bibinfo {pages} {722} (\bibinfo {year} {2002})}\BibitemShut {NoStop}%
	\bibitem [{\citenamefont {Hewson}(1997)}]{Hewson_Book}%
	\BibitemOpen
	\bibfield  {author} {\bibinfo {author} {\bibfnamefont {A.~C.}\ \bibnamefont
			{Hewson}},\ }\href@noop {} {\emph {\bibinfo {title} {The Kondo problem to
				heavy fermions}}}\ (\bibinfo  {publisher} {Cambr. Univ. Press},\ \bibinfo
	{address} {Cambridge},\ \bibinfo {year} {1997})\BibitemShut {NoStop}%
	\bibitem [{\citenamefont {Averin}\ and\ \citenamefont
		{Likharev}(1986)}]{Averin1986}%
	\BibitemOpen
	\bibfield  {author} {\bibinfo {author} {\bibfnamefont {D.~V.}\ \bibnamefont
			{Averin}}\ and\ \bibinfo {author} {\bibfnamefont {K.~K.}\ \bibnamefont
			{Likharev}},\ }\href {\doibase 10.1007/BF00683469} {\bibfield  {journal}
		{\bibinfo  {journal} {Journal of Low Temperature Physics}\ }\textbf {\bibinfo
			{volume} {62}},\ \bibinfo {pages} {345} (\bibinfo {year} {1986})}\BibitemShut
	{NoStop}%
	\bibitem [{\citenamefont {Shtrikman}\ \emph {et~al.}(1998)\citenamefont
		{Shtrikman}, \citenamefont {Mahalu}, \citenamefont {Abusch-Magder},
		\citenamefont {Meirav}, \citenamefont {Kastner},\ and\ \citenamefont
		{Goldhaber-Gordon}}]{Shtrikman1998}%
	\BibitemOpen
	\bibfield  {author} {\bibinfo {author} {\bibfnamefont {H.}~\bibnamefont
			{Shtrikman}}, \bibinfo {author} {\bibfnamefont {D.}~\bibnamefont {Mahalu}},
		\bibinfo {author} {\bibfnamefont {D.}~\bibnamefont {Abusch-Magder}}, \bibinfo
		{author} {\bibfnamefont {U.}~\bibnamefont {Meirav}}, \bibinfo {author}
		{\bibfnamefont {M.~A.}\ \bibnamefont {Kastner}}, \ and\ \bibinfo {author}
		{\bibfnamefont {D.}~\bibnamefont {Goldhaber-Gordon}},\ }\href@noop {}
	{\bibfield  {journal} {\bibinfo  {journal} {Nature}\ }\textbf {\bibinfo
			{volume} {391}},\ \bibinfo {pages} {156} (\bibinfo {year}
		{1998})}\BibitemShut {NoStop}%
	\bibitem [{\citenamefont {Goldhaber-Gordon}\ \emph {et~al.}(1998)\citenamefont
		{Goldhaber-Gordon}, \citenamefont {G\"ores}, \citenamefont {Kastner},
		\citenamefont {Shtrikman}, \citenamefont {Mahalu},\ and\ \citenamefont
		{Meirav}}]{Goldhaber1998}%
	\BibitemOpen
	\bibfield  {author} {\bibinfo {author} {\bibfnamefont {D.}~\bibnamefont
			{Goldhaber-Gordon}}, \bibinfo {author} {\bibfnamefont {J.}~\bibnamefont
			{G\"ores}}, \bibinfo {author} {\bibfnamefont {M.~A.}\ \bibnamefont
			{Kastner}}, \bibinfo {author} {\bibfnamefont {H.}~\bibnamefont {Shtrikman}},
		\bibinfo {author} {\bibfnamefont {D.}~\bibnamefont {Mahalu}}, \ and\ \bibinfo
		{author} {\bibfnamefont {U.}~\bibnamefont {Meirav}},\ }\href {\doibase
		10.1103/PhysRevLett.81.5225} {\bibfield  {journal} {\bibinfo  {journal}
			{Phys. Rev. Lett.}\ }\textbf {\bibinfo {volume} {81}},\ \bibinfo {pages}
		{5225} (\bibinfo {year} {1998})}\BibitemShut {NoStop}%
	\bibitem [{\citenamefont {Champagne}\ \emph {et~al.}(2005)\citenamefont
		{Champagne}, \citenamefont {Pasupathy},\ and\ \citenamefont
		{Ralph}}]{Champagne2005}%
	\BibitemOpen
	\bibfield  {author} {\bibinfo {author} {\bibfnamefont {A.~R.}\ \bibnamefont
			{Champagne}}, \bibinfo {author} {\bibfnamefont {A.~N.}\ \bibnamefont
			{Pasupathy}}, \ and\ \bibinfo {author} {\bibfnamefont {D.~C.}\ \bibnamefont
			{Ralph}},\ }\href@noop {} {\bibfield  {journal} {\bibinfo  {journal} {Nano
				Lett.}\ }\textbf {\bibinfo {volume} {5}},\ \bibinfo {pages} {305} (\bibinfo
		{year} {2005})}\BibitemShut {NoStop}%
	\bibitem [{\citenamefont {Parks}\ \emph {et~al.}(2007)\citenamefont {Parks},
		\citenamefont {Champagne}, \citenamefont {Hutchison}, \citenamefont
		{Flores-Torres}, \citenamefont {Abru\~na},\ and\ \citenamefont
		{Ralph}}]{Parks2007}%
	\BibitemOpen
	\bibfield  {author} {\bibinfo {author} {\bibfnamefont {J.~J.}\ \bibnamefont
			{Parks}}, \bibinfo {author} {\bibfnamefont {A.~R.}\ \bibnamefont
			{Champagne}}, \bibinfo {author} {\bibfnamefont {G.~R.}\ \bibnamefont
			{Hutchison}}, \bibinfo {author} {\bibfnamefont {S.}~\bibnamefont
			{Flores-Torres}}, \bibinfo {author} {\bibfnamefont {H.~D.}\ \bibnamefont
			{Abru\~na}}, \ and\ \bibinfo {author} {\bibfnamefont {D.~C.}\ \bibnamefont
			{Ralph}},\ }\href {\doibase 10.1103/PhysRevLett.99.026601} {\bibfield
		{journal} {\bibinfo  {journal} {Phys. Rev. Lett.}\ }\textbf {\bibinfo
			{volume} {99}},\ \bibinfo {pages} {026601} (\bibinfo {year}
		{2007})}\BibitemShut {NoStop}%
	\bibitem [{\citenamefont {Jarillo-Herrero}\ \emph {et~al.}(2005)\citenamefont
		{Jarillo-Herrero}, \citenamefont {Kong}, \citenamefont {van~der Zant},
		\citenamefont {Dekker}, \citenamefont {Kouwenhoven},\ and\ \citenamefont
		{DeFranceschi}}]{Jarillo2005}%
	\BibitemOpen
	\bibfield  {author} {\bibinfo {author} {\bibfnamefont {P.}~\bibnamefont
			{Jarillo-Herrero}}, \bibinfo {author} {\bibfnamefont {J.}~\bibnamefont
			{Kong}}, \bibinfo {author} {\bibfnamefont {H.~S.~J.}\ \bibnamefont {van~der
				Zant}}, \bibinfo {author} {\bibfnamefont {C.}~\bibnamefont {Dekker}},
		\bibinfo {author} {\bibfnamefont {L.~P.}\ \bibnamefont {Kouwenhoven}}, \ and\
		\bibinfo {author} {\bibfnamefont {S.}~\bibnamefont {DeFranceschi}},\
	}\href@noop {} {\bibfield  {journal} {\bibinfo  {journal} {Nature}\ }\textbf
		{\bibinfo {volume} {434}},\ \bibinfo {pages} {484} (\bibinfo {year}
		{2005})}\BibitemShut {NoStop}%
	\bibitem [{\citenamefont {Roch}\ \emph {et~al.}(2009)\citenamefont {Roch},
		\citenamefont {Florens}, \citenamefont {Costi}, \citenamefont {Wernsdorfer},\
		and\ \citenamefont {Balestro}}]{Roch2009}%
	\BibitemOpen
	\bibfield  {author} {\bibinfo {author} {\bibfnamefont {N.}~\bibnamefont
			{Roch}}, \bibinfo {author} {\bibfnamefont {S.}~\bibnamefont {Florens}},
		\bibinfo {author} {\bibfnamefont {T.~A.}\ \bibnamefont {Costi}}, \bibinfo
		{author} {\bibfnamefont {W.}~\bibnamefont {Wernsdorfer}}, \ and\ \bibinfo
		{author} {\bibfnamefont {F.}~\bibnamefont {Balestro}},\ }\href {\doibase
		10.1103/PhysRevLett.103.197202} {\bibfield  {journal} {\bibinfo  {journal}
			{Phys. Rev. Lett.}\ }\textbf {\bibinfo {volume} {103}},\ \bibinfo {pages}
		{197202} (\bibinfo {year} {2009})}\BibitemShut {NoStop}%
	\bibitem [{\citenamefont {Parks}\ \emph {et~al.}(2010)\citenamefont {Parks},
		\citenamefont {Champagne}, \citenamefont {Costi}, \citenamefont {Shum},
		\citenamefont {Pasupathy}, \citenamefont {Neuscamman}, \citenamefont
		{Flores-Torres}, \citenamefont {Cornaglia}, \citenamefont {Aligia},
		\citenamefont {Balseiro}, \citenamefont {Chan}, \citenamefont {Abru\~na},\
		and\ \citenamefont {Ralph}}]{Parks2010}%
	\BibitemOpen
	\bibfield  {author} {\bibinfo {author} {\bibfnamefont {J.~J.}\ \bibnamefont
			{Parks}}, \bibinfo {author} {\bibfnamefont {A.~R.}\ \bibnamefont
			{Champagne}}, \bibinfo {author} {\bibfnamefont {T.~A.}\ \bibnamefont
			{Costi}}, \bibinfo {author} {\bibfnamefont {W.~W.}\ \bibnamefont {Shum}},
		\bibinfo {author} {\bibfnamefont {A.~N.}\ \bibnamefont {Pasupathy}}, \bibinfo
		{author} {\bibfnamefont {E.}~\bibnamefont {Neuscamman}}, \bibinfo {author}
		{\bibfnamefont {S.}~\bibnamefont {Flores-Torres}}, \bibinfo {author}
		{\bibfnamefont {P.~S.}\ \bibnamefont {Cornaglia}}, \bibinfo {author}
		{\bibfnamefont {A.~A.}\ \bibnamefont {Aligia}}, \bibinfo {author}
		{\bibfnamefont {C.~A.}\ \bibnamefont {Balseiro}}, \bibinfo {author}
		{\bibfnamefont {G.~K.-L.}\ \bibnamefont {Chan}}, \bibinfo {author}
		{\bibfnamefont {H.~D.}\ \bibnamefont {Abru\~na}}, \ and\ \bibinfo {author}
		{\bibfnamefont {D.~C.}\ \bibnamefont {Ralph}},\ }\href {\doibase
		10.1126/science.1186874} {\bibfield  {journal} {\bibinfo  {journal}
			{Science}\ }\textbf {\bibinfo {volume} {328}},\ \bibinfo {pages} {1370}
		(\bibinfo {year} {2010})}\BibitemShut {NoStop}%
	\bibitem [{\citenamefont {Avinun-Kalish}\ \emph {et~al.}(2005)\citenamefont
		{Avinun-Kalish}, \citenamefont {Heiblum}, \citenamefont {Zarchin},
		\citenamefont {Mahalu},\ and\ \citenamefont {Umansky}}]{avinun2005crossover}%
	\BibitemOpen
	\bibfield  {author} {\bibinfo {author} {\bibfnamefont {M.}~\bibnamefont
			{Avinun-Kalish}}, \bibinfo {author} {\bibfnamefont {M.}~\bibnamefont
			{Heiblum}}, \bibinfo {author} {\bibfnamefont {O.}~\bibnamefont {Zarchin}},
		\bibinfo {author} {\bibfnamefont {D.}~\bibnamefont {Mahalu}}, \ and\ \bibinfo
		{author} {\bibfnamefont {V.}~\bibnamefont {Umansky}},\ }\href@noop {}
	{\bibfield  {journal} {\bibinfo  {journal} {Nature}\ }\textbf {\bibinfo
			{volume} {436}},\ \bibinfo {pages} {529} (\bibinfo {year}
		{2005})}\BibitemShut {NoStop}%
	\bibitem [{\citenamefont {Yacoby}\ \emph {et~al.}(1995)\citenamefont {Yacoby},
		\citenamefont {Heiblum}, \citenamefont {Mahalu},\ and\ \citenamefont
		{Shtrikman}}]{yacoby1995coherence}%
	\BibitemOpen
	\bibfield  {author} {\bibinfo {author} {\bibfnamefont {A.}~\bibnamefont
			{Yacoby}}, \bibinfo {author} {\bibfnamefont {M.}~\bibnamefont {Heiblum}},
		\bibinfo {author} {\bibfnamefont {D.}~\bibnamefont {Mahalu}}, \ and\ \bibinfo
		{author} {\bibfnamefont {H.}~\bibnamefont {Shtrikman}},\ }\href@noop {}
	{\bibfield  {journal} {\bibinfo  {journal} {Phys. Rev. Lett.}\ }\textbf
		{\bibinfo {volume} {74}},\ \bibinfo {pages} {4047} (\bibinfo {year}
		{1995})}\BibitemShut {NoStop}%
	\bibitem [{\citenamefont {Schuster}\ \emph {et~al.}(1997)\citenamefont
		{Schuster}, \citenamefont {Buks}, \citenamefont {Heiblum}, \citenamefont
		{Mahalu}, \citenamefont {Umansky},\ and\ \citenamefont
		{Shtrikman}}]{schuster1997phase}%
	\BibitemOpen
	\bibfield  {author} {\bibinfo {author} {\bibfnamefont {R.}~\bibnamefont
			{Schuster}}, \bibinfo {author} {\bibfnamefont {E.}~\bibnamefont {Buks}},
		\bibinfo {author} {\bibfnamefont {M.}~\bibnamefont {Heiblum}}, \bibinfo
		{author} {\bibfnamefont {D.}~\bibnamefont {Mahalu}}, \bibinfo {author}
		{\bibfnamefont {V.}~\bibnamefont {Umansky}}, \ and\ \bibinfo {author}
		{\bibfnamefont {H.}~\bibnamefont {Shtrikman}},\ }\href@noop {} {\bibfield
		{journal} {\bibinfo  {journal} {Nature}\ }\textbf {\bibinfo {volume} {385}},\
		\bibinfo {pages} {417} (\bibinfo {year} {1997})}\BibitemShut {NoStop}%
	\bibitem [{\citenamefont {Aikawa}\ \emph {et~al.}(2004)\citenamefont {Aikawa},
		\citenamefont {Kobayashi}, \citenamefont {Sano}, \citenamefont {Katsumoto},\
		and\ \citenamefont {Iye}}]{aikawa2004interference}%
	\BibitemOpen
	\bibfield  {author} {\bibinfo {author} {\bibfnamefont {H.}~\bibnamefont
			{Aikawa}}, \bibinfo {author} {\bibfnamefont {K.}~\bibnamefont {Kobayashi}},
		\bibinfo {author} {\bibfnamefont {A.}~\bibnamefont {Sano}}, \bibinfo {author}
		{\bibfnamefont {S.}~\bibnamefont {Katsumoto}}, \ and\ \bibinfo {author}
		{\bibfnamefont {Y.}~\bibnamefont {Iye}},\ }\href@noop {} {\bibfield
		{journal} {\bibinfo  {journal} {J. Phys. Soc. Jpn.}\ }\textbf {\bibinfo
			{volume} {73}},\ \bibinfo {pages} {3235} (\bibinfo {year}
		{2004})}\BibitemShut {NoStop}%
	\bibitem [{\citenamefont {Ji}\ \emph {et~al.}(2002)\citenamefont {Ji},
		\citenamefont {Heiblum},\ and\ \citenamefont
		{Shtrikman}}]{ji2002transmission}%
	\BibitemOpen
	\bibfield  {author} {\bibinfo {author} {\bibfnamefont {Y.}~\bibnamefont
			{Ji}}, \bibinfo {author} {\bibfnamefont {M.}~\bibnamefont {Heiblum}}, \ and\
		\bibinfo {author} {\bibfnamefont {H.}~\bibnamefont {Shtrikman}},\ }\href@noop
	{} {\bibfield  {journal} {\bibinfo  {journal} {Phys. Rev. Lett.}\ }\textbf
		{\bibinfo {volume} {88}},\ \bibinfo {pages} {076601} (\bibinfo {year}
		{2002})}\BibitemShut {NoStop}%
	\bibitem [{\citenamefont {Zaffalon}\ \emph {et~al.}(2008)\citenamefont
		{Zaffalon}, \citenamefont {Bid}, \citenamefont {Heiblum}, \citenamefont
		{Mahalu},\ and\ \citenamefont {Umansky}}]{zaffalon2008transmission}%
	\BibitemOpen
	\bibfield  {author} {\bibinfo {author} {\bibfnamefont {M.}~\bibnamefont
			{Zaffalon}}, \bibinfo {author} {\bibfnamefont {A.}~\bibnamefont {Bid}},
		\bibinfo {author} {\bibfnamefont {M.}~\bibnamefont {Heiblum}}, \bibinfo
		{author} {\bibfnamefont {D.}~\bibnamefont {Mahalu}}, \ and\ \bibinfo {author}
		{\bibfnamefont {V.}~\bibnamefont {Umansky}},\ }\href@noop {} {\bibfield
		{journal} {\bibinfo  {journal} {Phys. Rev. Lett.}\ }\textbf {\bibinfo
			{volume} {100}},\ \bibinfo {pages} {226601} (\bibinfo {year}
		{2008})}\BibitemShut {NoStop}%
	\bibitem [{\citenamefont {Lim}\ \emph {et~al.}(2006)\citenamefont {Lim},
		\citenamefont {Choi}, \citenamefont {Choi}, \citenamefont {L{\'o}pez},\ and\
		\citenamefont {Aguado}}]{lim2006kondo}%
	\BibitemOpen
	\bibfield  {author} {\bibinfo {author} {\bibfnamefont {J.~S.}\ \bibnamefont
			{Lim}}, \bibinfo {author} {\bibfnamefont {M.-S.}\ \bibnamefont {Choi}},
		\bibinfo {author} {\bibfnamefont {M.}~\bibnamefont {Choi}}, \bibinfo {author}
		{\bibfnamefont {R.}~\bibnamefont {L{\'o}pez}}, \ and\ \bibinfo {author}
		{\bibfnamefont {R.}~\bibnamefont {Aguado}},\ }\href@noop {} {\bibfield
		{journal} {\bibinfo  {journal} {Phys. Rev. B}\ }\textbf {\bibinfo {volume}
			{74}},\ \bibinfo {pages} {205119} (\bibinfo {year} {2006})}\BibitemShut
	{NoStop}%
	\bibitem [{\citenamefont {Kleeorin}\ and\ \citenamefont
		{Meir}(2017{\natexlab{a}})}]{kleeorin2017abrupt}%
	\BibitemOpen
	\bibfield  {author} {\bibinfo {author} {\bibfnamefont {Y.}~\bibnamefont
			{Kleeorin}}\ and\ \bibinfo {author} {\bibfnamefont {Y.}~\bibnamefont
			{Meir}},\ }\href@noop {} {\bibfield  {journal} {\bibinfo  {journal} {Phys.
				Rev. B}\ }\textbf {\bibinfo {volume} {96}},\ \bibinfo {pages} {045118}
		(\bibinfo {year} {2017}{\natexlab{a}})}\BibitemShut {NoStop}%
	\bibitem [{\citenamefont {B{\"u}sser}\ \emph {et~al.}(2011)\citenamefont
		{B{\"u}sser}, \citenamefont {Vernek}, \citenamefont {Orellana}, \citenamefont
		{Lara}, \citenamefont {Kim}, \citenamefont {Feiguin}, \citenamefont {Anda},\
		and\ \citenamefont {Martins}}]{busser2011transport}%
	\BibitemOpen
	\bibfield  {author} {\bibinfo {author} {\bibfnamefont {C.}~\bibnamefont
			{B{\"u}sser}}, \bibinfo {author} {\bibfnamefont {E.}~\bibnamefont {Vernek}},
		\bibinfo {author} {\bibfnamefont {P.}~\bibnamefont {Orellana}}, \bibinfo
		{author} {\bibfnamefont {G.}~\bibnamefont {Lara}}, \bibinfo {author}
		{\bibfnamefont {E.}~\bibnamefont {Kim}}, \bibinfo {author} {\bibfnamefont
			{A.}~\bibnamefont {Feiguin}}, \bibinfo {author} {\bibfnamefont
			{E.}~\bibnamefont {Anda}}, \ and\ \bibinfo {author} {\bibfnamefont
			{G.}~\bibnamefont {Martins}},\ }\href@noop {} {\bibfield  {journal} {\bibinfo
			{journal} {Phys. Rev. B}\ }\textbf {\bibinfo {volume} {83}},\ \bibinfo
		{pages} {125404} (\bibinfo {year} {2011})}\BibitemShut {NoStop}%
	\bibitem [{\citenamefont {Roura-Bas}\ \emph {et~al.}(2011)\citenamefont
		{Roura-Bas}, \citenamefont {Tosi}, \citenamefont {Aligia},\ and\
		\citenamefont {Hallberg}}]{roura2011interplay}%
	\BibitemOpen
	\bibfield  {author} {\bibinfo {author} {\bibfnamefont {P.}~\bibnamefont
			{Roura-Bas}}, \bibinfo {author} {\bibfnamefont {L.}~\bibnamefont {Tosi}},
		\bibinfo {author} {\bibfnamefont {A.}~\bibnamefont {Aligia}}, \ and\ \bibinfo
		{author} {\bibfnamefont {K.}~\bibnamefont {Hallberg}},\ }\href@noop {}
	{\bibfield  {journal} {\bibinfo  {journal} {Phys. Rev. B}\ }\textbf {\bibinfo
			{volume} {84}},\ \bibinfo {pages} {073406} (\bibinfo {year}
		{2011})}\BibitemShut {NoStop}%
	\bibitem [{\citenamefont {Weymann}\ \emph {et~al.}(2018)\citenamefont
		{Weymann}, \citenamefont {Chirla}, \citenamefont {Trocha},\ and\
		\citenamefont {Moca}}]{weymann20184}%
	\BibitemOpen
	\bibfield  {author} {\bibinfo {author} {\bibfnamefont {I.}~\bibnamefont
			{Weymann}}, \bibinfo {author} {\bibfnamefont {R.}~\bibnamefont {Chirla}},
		\bibinfo {author} {\bibfnamefont {P.}~\bibnamefont {Trocha}}, \ and\ \bibinfo
		{author} {\bibfnamefont {C.~P.}\ \bibnamefont {Moca}},\ }\href@noop {}
	{\bibfield  {journal} {\bibinfo  {journal} {Phys. Rev. B}\ }\textbf {\bibinfo
			{volume} {97}},\ \bibinfo {pages} {085404} (\bibinfo {year}
		{2018})}\BibitemShut {NoStop}%
	\bibitem [{\citenamefont {Silvestrov}\ and\ \citenamefont
		{Imry}(2007)}]{silvestrov2007level}%
	\BibitemOpen
	\bibfield  {author} {\bibinfo {author} {\bibfnamefont {P.}~\bibnamefont
			{Silvestrov}}\ and\ \bibinfo {author} {\bibfnamefont {Y.}~\bibnamefont
			{Imry}},\ }\href@noop {} {\bibfield  {journal} {\bibinfo  {journal} {New J.
				Phys.}\ }\textbf {\bibinfo {volume} {9}},\ \bibinfo {pages} {125} (\bibinfo
		{year} {2007})}\BibitemShut {NoStop}%
	\bibitem [{\citenamefont {Hohenberg}\ and\ \citenamefont
		{Kohn}(1964)}]{HohenbergKohn:64}%
	\BibitemOpen
	\bibfield  {author} {\bibinfo {author} {\bibfnamefont {P.}~\bibnamefont
			{Hohenberg}}\ and\ \bibinfo {author} {\bibfnamefont {W.}~\bibnamefont
			{Kohn}},\ }\href@noop {} {\bibfield  {journal} {\bibinfo  {journal} {Phys.
				Rev}\ }\textbf {\bibinfo {volume} {136}},\ \bibinfo {pages} {B864} (\bibinfo
		{year} {1964})}\BibitemShut {NoStop}%
	\bibitem [{\citenamefont {Kohn}\ and\ \citenamefont
		{Sham}(1965)}]{Kohn:PR:1965}%
	\BibitemOpen
	\bibfield  {author} {\bibinfo {author} {\bibfnamefont {W.}~\bibnamefont
			{Kohn}}\ and\ \bibinfo {author} {\bibfnamefont {L.~J.}\ \bibnamefont
			{Sham}},\ }\href {\doibase 10.1103/PhysRev.140.A1133} {\bibfield  {journal}
		{\bibinfo  {journal} {Phys. Rev.}\ }\textbf {\bibinfo {volume} {140}},\
		\bibinfo {pages} {A1133} (\bibinfo {year} {1965})}\BibitemShut {NoStop}%
	\bibitem [{\citenamefont {Dreizler}\ and\ \citenamefont
		{Gross}(1990)}]{DreizlerGross:90}%
	\BibitemOpen
	\bibfield  {author} {\bibinfo {author} {\bibfnamefont {R.~M.}\ \bibnamefont
			{Dreizler}}\ and\ \bibinfo {author} {\bibfnamefont {E.~K.~U.}\ \bibnamefont
			{Gross}},\ }\href@noop {} {\emph {\bibinfo {title} {Density Functional
				Theory}}}\ (\bibinfo  {publisher} {Springer},\ \bibinfo {address} {Berlin},\
	\bibinfo {year} {1990})\BibitemShut {NoStop}%
	\bibitem [{\citenamefont {{J.P.~Perdew}}(1985)}]{Perdew:85}%
	\BibitemOpen
	\bibfield  {author} {\bibinfo {author} {\bibnamefont {{J.P.~Perdew}}},\
	}\href@noop {} {\bibfield  {journal} {\bibinfo  {journal} {Phys. Rev. Lett.}\
		}\textbf {\bibinfo {volume} {55}},\ \bibinfo {pages} {1665} (\bibinfo {year}
		{1985})}\BibitemShut {NoStop}%
	\bibitem [{\citenamefont {Becke}(1988)}]{becke1988_gga}%
	\BibitemOpen
	\bibfield  {author} {\bibinfo {author} {\bibfnamefont {A.~D.}\ \bibnamefont
			{Becke}},\ }\href {\doibase 10.1103/PhysRevA.38.3098} {\bibfield  {journal}
		{\bibinfo  {journal} {Phys. Rev. A}\ }\textbf {\bibinfo {volume} {38}},\
		\bibinfo {pages} {3098} (\bibinfo {year} {1988})}\BibitemShut {NoStop}%
	\bibitem [{\citenamefont {Perdew}\ \emph {et~al.}(1996)\citenamefont {Perdew},
		\citenamefont {Burke},\ and\ \citenamefont
		{Ernzerhof}}]{perdew1996generalized}%
	\BibitemOpen
	\bibfield  {author} {\bibinfo {author} {\bibfnamefont {J.~P.}\ \bibnamefont
			{Perdew}}, \bibinfo {author} {\bibfnamefont {K.}~\bibnamefont {Burke}}, \
		and\ \bibinfo {author} {\bibfnamefont {M.}~\bibnamefont {Ernzerhof}},\
	}\href@noop {} {\bibfield  {journal} {\bibinfo  {journal} {Phys. Rev. Lett.}\
		}\textbf {\bibinfo {volume} {77}},\ \bibinfo {pages} {3865} (\bibinfo {year}
		{1996})}\BibitemShut {NoStop}%
	\bibitem [{\citenamefont {{A.D.~Becke}}(1993)}]{Becke:93-2}%
	\BibitemOpen
	\bibfield  {author} {\bibinfo {author} {\bibnamefont {{A.D.~Becke}}},\
	}\href@noop {} {\bibfield  {journal} {\bibinfo  {journal} {J. Chem. Phys.}\
		}\textbf {\bibinfo {volume} {98}},\ \bibinfo {pages} {5648} (\bibinfo {year}
		{1993})}\BibitemShut {NoStop}%
	\bibitem [{\citenamefont {Stefanucci}\ and\ \citenamefont
		{Kurth}(2011)}]{sk.2011}%
	\BibitemOpen
	\bibfield  {author} {\bibinfo {author} {\bibfnamefont {G.}~\bibnamefont
			{Stefanucci}}\ and\ \bibinfo {author} {\bibfnamefont {S.}~\bibnamefont
			{Kurth}},\ }\href {\doibase 10.1103/PhysRevLett.107.216401} {\bibfield
		{journal} {\bibinfo  {journal} {Phys. Rev. Lett.}\ }\textbf {\bibinfo
			{volume} {107}},\ \bibinfo {pages} {216401} (\bibinfo {year}
		{2011})}\BibitemShut {NoStop}%
	\bibitem [{\citenamefont {Bergfield}\ \emph {et~al.}(2012)\citenamefont
		{Bergfield}, \citenamefont {Liu}, \citenamefont {Burke},\ and\ \citenamefont
		{Stafford}}]{blbs.2012}%
	\BibitemOpen
	\bibfield  {author} {\bibinfo {author} {\bibfnamefont {J.~P.}\ \bibnamefont
			{Bergfield}}, \bibinfo {author} {\bibfnamefont {Z.-F.}\ \bibnamefont {Liu}},
		\bibinfo {author} {\bibfnamefont {K.}~\bibnamefont {Burke}}, \ and\ \bibinfo
		{author} {\bibfnamefont {C.~A.}\ \bibnamefont {Stafford}},\ }\href {\doibase
		10.1103/PhysRevLett.108.066801} {\bibfield  {journal} {\bibinfo  {journal}
			{Phys. Rev. Lett.}\ }\textbf {\bibinfo {volume} {108}},\ \bibinfo {pages}
		{066801} (\bibinfo {year} {2012})}\BibitemShut {NoStop}%
	\bibitem [{\citenamefont {Tr\"oster}\ \emph {et~al.}(2012)\citenamefont
		{Tr\"oster}, \citenamefont {Schmitteckert},\ and\ \citenamefont
		{Evers}}]{tse.2012}%
	\BibitemOpen
	\bibfield  {author} {\bibinfo {author} {\bibfnamefont {P.}~\bibnamefont
			{Tr\"oster}}, \bibinfo {author} {\bibfnamefont {P.}~\bibnamefont
			{Schmitteckert}}, \ and\ \bibinfo {author} {\bibfnamefont {F.}~\bibnamefont
			{Evers}},\ }\href {\doibase 10.1103/PhysRevB.85.115409} {\bibfield  {journal}
		{\bibinfo  {journal} {Phys. Rev. B}\ }\textbf {\bibinfo {volume} {85}},\
		\bibinfo {pages} {115409} (\bibinfo {year} {2012})}\BibitemShut {NoStop}%
	\bibitem [{\citenamefont {Lima}\ \emph {et~al.}(2002)\citenamefont {Lima},
		\citenamefont {Oliveira},\ and\ \citenamefont
		{Capelle}}]{LimaOliveiraCapelle:02}%
	\BibitemOpen
	\bibfield  {author} {\bibinfo {author} {\bibfnamefont {N.~A.}\ \bibnamefont
			{Lima}}, \bibinfo {author} {\bibfnamefont {L.~N.}\ \bibnamefont {Oliveira}},
		\ and\ \bibinfo {author} {\bibfnamefont {K.}~\bibnamefont {Capelle}},\ }\href
	{\doibase 10.1209/epl/i2002-00261-y} {\bibfield  {journal} {\bibinfo
			{journal} {Europhys. Lett.}\ }\textbf {\bibinfo {volume} {60}},\ \bibinfo
		{pages} {601} (\bibinfo {year} {2002})}\BibitemShut {NoStop}%
	\bibitem [{\citenamefont {Jacob}\ and\ \citenamefont
		{Kurth}(2018)}]{JacobKurth:18}%
	\BibitemOpen
	\bibfield  {author} {\bibinfo {author} {\bibfnamefont {D.}~\bibnamefont
			{Jacob}}\ and\ \bibinfo {author} {\bibfnamefont {S.}~\bibnamefont {Kurth}},\
	}\href@noop {} {\bibfield  {journal} {\bibinfo  {journal} {Nano Lett.}\
		}\textbf {\bibinfo {volume} {18}},\ \bibinfo {pages} {2086} (\bibinfo {year}
		{2018})}\BibitemShut {NoStop}%
	\bibitem [{\citenamefont {Kurth}\ \emph {et~al.}(2019)\citenamefont {Kurth},
		\citenamefont {Jacob}, \citenamefont {Sobrino},\ and\ \citenamefont
		{Stefanucci}}]{kurth2019nonequilibrium}%
	\BibitemOpen
	\bibfield  {author} {\bibinfo {author} {\bibfnamefont {S.}~\bibnamefont
			{Kurth}}, \bibinfo {author} {\bibfnamefont {D.}~\bibnamefont {Jacob}},
		\bibinfo {author} {\bibfnamefont {N.}~\bibnamefont {Sobrino}}, \ and\
		\bibinfo {author} {\bibfnamefont {G.}~\bibnamefont {Stefanucci}},\
	}\href@noop {} {\bibfield  {journal} {\bibinfo  {journal} {Phys. Rev. B}\
		}\textbf {\bibinfo {volume} {100}},\ \bibinfo {pages} {085114} (\bibinfo
		{year} {2019})}\BibitemShut {NoStop}%
	\bibitem [{\citenamefont {Jacob}\ \emph {et~al.}(2020)\citenamefont {Jacob},
		\citenamefont {Stefanucci},\ and\ \citenamefont
		{Kurth}}]{JacobStefanucciKurth:20}%
	\BibitemOpen
	\bibfield  {author} {\bibinfo {author} {\bibfnamefont {D.}~\bibnamefont
			{Jacob}}, \bibinfo {author} {\bibfnamefont {G.}~\bibnamefont {Stefanucci}}, \
		and\ \bibinfo {author} {\bibfnamefont {S.}~\bibnamefont {Kurth}},\ }\href
	{\doibase 10.1103/PhysRevLett.125.216401} {\bibfield  {journal} {\bibinfo
			{journal} {Phys. Rev. Lett.}\ }\textbf {\bibinfo {volume} {125}},\ \bibinfo
		{pages} {216401} (\bibinfo {year} {2020})}\BibitemShut {NoStop}%
	\bibitem [{\citenamefont {Sobrino}\ \emph {et~al.}(2020)\citenamefont
		{Sobrino}, \citenamefont {Kurth},\ and\ \citenamefont
		{Jacob}}]{sobrino2020exchange}%
	\BibitemOpen
	\bibfield  {author} {\bibinfo {author} {\bibfnamefont {N.}~\bibnamefont
			{Sobrino}}, \bibinfo {author} {\bibfnamefont {S.}~\bibnamefont {Kurth}}, \
		and\ \bibinfo {author} {\bibfnamefont {D.}~\bibnamefont {Jacob}},\
	}\href@noop {} {\bibfield  {journal} {\bibinfo  {journal} {Phys. Rev. B}\
		}\textbf {\bibinfo {volume} {102}},\ \bibinfo {pages} {035159} (\bibinfo
		{year} {2020})}\BibitemShut {NoStop}%
	\bibitem [{\citenamefont {Perdew}\ \emph {et~al.}(1982)\citenamefont {Perdew},
		\citenamefont {Parr}, \citenamefont {Levy},\ and\ \citenamefont
		{Balduz}}]{PerdewParrLevyBalduz:82}%
	\BibitemOpen
	\bibfield  {author} {\bibinfo {author} {\bibfnamefont {J.~P.}\ \bibnamefont
			{Perdew}}, \bibinfo {author} {\bibfnamefont {R.}~\bibnamefont {Parr}},
		\bibinfo {author} {\bibfnamefont {M.}~\bibnamefont {Levy}}, \ and\ \bibinfo
		{author} {\bibfnamefont {J.~L.}\ \bibnamefont {Balduz}},\ }\href@noop {}
	{\bibfield  {journal} {\bibinfo  {journal} {Phys. Rev. Lett.}\ }\textbf
		{\bibinfo {volume} {49}},\ \bibinfo {pages} {1691} (\bibinfo {year}
		{1982})}\BibitemShut {NoStop}%
	\bibitem [{\citenamefont {Kurth}\ and\ \citenamefont
		{Stefanucci}(2016{\natexlab{a}})}]{KurthStefanucci:16}%
	\BibitemOpen
	\bibfield  {author} {\bibinfo {author} {\bibfnamefont {S.}~\bibnamefont
			{Kurth}}\ and\ \bibinfo {author} {\bibfnamefont {G.}~\bibnamefont
			{Stefanucci}},\ }\href {\doibase 10.1103/PhysRevB.94.241103} {\bibfield
		{journal} {\bibinfo  {journal} {Phys. Rev. B}\ }\textbf {\bibinfo {volume}
			{94}},\ \bibinfo {pages} {241103(R)} (\bibinfo {year}
		{2016}{\natexlab{a}})}\BibitemShut {NoStop}%
	\bibitem [{\citenamefont {Kurth}\ and\ \citenamefont
		{Stefanucci}(2017)}]{kurth2017transport}%
	\BibitemOpen
	\bibfield  {author} {\bibinfo {author} {\bibfnamefont {S.}~\bibnamefont
			{Kurth}}\ and\ \bibinfo {author} {\bibfnamefont {G.}~\bibnamefont
			{Stefanucci}},\ }\href@noop {} {\bibfield  {journal} {\bibinfo  {journal} {J.
				Phys.: Condens. Matter}\ }\textbf {\bibinfo {volume} {29}},\ \bibinfo {pages}
		{413002} (\bibinfo {year} {2017})}\BibitemShut {NoStop}%
	\bibitem [{Note1()}]{Note1}%
	\BibitemOpen
	\bibinfo {note} {See, e.g., Ch.~5.2 in the book by Hewson\cite
		{Hewson_Book}}\BibitemShut {NoStop}%
	\bibitem [{\citenamefont {Mera}\ \emph {et~al.}(2010)\citenamefont {Mera},
		\citenamefont {Kaasbjerg}, \citenamefont {Niquet},\ and\ \citenamefont
		{Stefanucci}}]{mera2010assessing}%
	\BibitemOpen
	\bibfield  {author} {\bibinfo {author} {\bibfnamefont {H.}~\bibnamefont
			{Mera}}, \bibinfo {author} {\bibfnamefont {K.}~\bibnamefont {Kaasbjerg}},
		\bibinfo {author} {\bibfnamefont {Y.}~\bibnamefont {Niquet}}, \ and\ \bibinfo
		{author} {\bibfnamefont {G.}~\bibnamefont {Stefanucci}},\ }\href@noop {}
	{\bibfield  {journal} {\bibinfo  {journal} {Phys. Rev. B}\ }\textbf {\bibinfo
			{volume} {81}},\ \bibinfo {pages} {035110} (\bibinfo {year}
		{2010})}\BibitemShut {NoStop}%
	\bibitem [{\citenamefont {Langreth}(1966)}]{langreth1966friedel}%
	\BibitemOpen
	\bibfield  {author} {\bibinfo {author} {\bibfnamefont {D.~C.}\ \bibnamefont
			{Langreth}},\ }\href@noop {} {\bibfield  {journal} {\bibinfo  {journal}
			{Phys. Rev.}\ }\textbf {\bibinfo {volume} {150}},\ \bibinfo {pages} {516}
		(\bibinfo {year} {1966})}\BibitemShut {NoStop}%
	\bibitem [{\citenamefont {Mera}\ and\ \citenamefont
		{Niquet}(2010)}]{mera2010kohn}%
	\BibitemOpen
	\bibfield  {author} {\bibinfo {author} {\bibfnamefont {H.}~\bibnamefont
			{Mera}}\ and\ \bibinfo {author} {\bibfnamefont {Y.}~\bibnamefont {Niquet}},\
	}\href@noop {} {\bibfield  {journal} {\bibinfo  {journal} {Phys. Rev. Lett.}\
		}\textbf {\bibinfo {volume} {105}},\ \bibinfo {pages} {216408} (\bibinfo
		{year} {2010})}\BibitemShut {NoStop}%
	\bibitem [{\citenamefont {Stefanucci}\ and\ \citenamefont
		{Kurth}(2015)}]{StefanucciKurth:15}%
	\BibitemOpen
	\bibfield  {author} {\bibinfo {author} {\bibfnamefont {G.}~\bibnamefont
			{Stefanucci}}\ and\ \bibinfo {author} {\bibfnamefont {S.}~\bibnamefont
			{Kurth}},\ }\href {\doibase DOI: 10.1021/acs.nanolett.5b03294} {\bibfield
		{journal} {\bibinfo  {journal} {Nano~Lett.}\ }\textbf {\bibinfo {volume}
			{15}},\ \bibinfo {pages} {8020} (\bibinfo {year} {2015})}\BibitemShut
	{NoStop}%
	\bibitem [{\citenamefont {Sobrino}\ \emph {et~al.}(2021)\citenamefont
		{Sobrino}, \citenamefont {Eich}, \citenamefont {Stefanucci}, \citenamefont
		{D'Agosta},\ and\ \citenamefont {Kurth}}]{sobrino2021thermoelectric}%
	\BibitemOpen
	\bibfield  {author} {\bibinfo {author} {\bibfnamefont {N.}~\bibnamefont
			{Sobrino}}, \bibinfo {author} {\bibfnamefont {F.}~\bibnamefont {Eich}},
		\bibinfo {author} {\bibfnamefont {G.}~\bibnamefont {Stefanucci}}, \bibinfo
		{author} {\bibfnamefont {R.}~\bibnamefont {D'Agosta}}, \ and\ \bibinfo
		{author} {\bibfnamefont {S.}~\bibnamefont {Kurth}},\ }\href@noop {}
	{\bibfield  {journal} {\bibinfo  {journal} {Phys. Rev. B}\ }\textbf {\bibinfo
			{volume} {104}},\ \bibinfo {pages} {125115} (\bibinfo {year}
		{2021})}\BibitemShut {NoStop}%
	\bibitem [{\citenamefont {Sobrino}\ \emph {et~al.}(2019)\citenamefont
		{Sobrino}, \citenamefont {D'Agosta},\ and\ \citenamefont
		{Kurth}}]{sobrino2019steady}%
	\BibitemOpen
	\bibfield  {author} {\bibinfo {author} {\bibfnamefont {N.}~\bibnamefont
			{Sobrino}}, \bibinfo {author} {\bibfnamefont {R.}~\bibnamefont {D'Agosta}}, \
		and\ \bibinfo {author} {\bibfnamefont {S.}~\bibnamefont {Kurth}},\
	}\href@noop {} {\bibfield  {journal} {\bibinfo  {journal} {Phys. Rev. B}\
		}\textbf {\bibinfo {volume} {100}},\ \bibinfo {pages} {195142} (\bibinfo
		{year} {2019})}\BibitemShut {NoStop}%
	\bibitem [{\citenamefont {Kurth}\ and\ \citenamefont
		{Stefanucci}(2016{\natexlab{b}})}]{kurth2016nonequilibrium}%
	\BibitemOpen
	\bibfield  {author} {\bibinfo {author} {\bibfnamefont {S.}~\bibnamefont
			{Kurth}}\ and\ \bibinfo {author} {\bibfnamefont {G.}~\bibnamefont
			{Stefanucci}},\ }\href@noop {} {\bibfield  {journal} {\bibinfo  {journal}
			{Phys. Rev. B}\ }\textbf {\bibinfo {volume} {94}},\ \bibinfo {pages}
		{241103(R)} (\bibinfo {year} {2016}{\natexlab{b}})}\BibitemShut {NoStop}%
	\bibitem [{\citenamefont {Stefanucci}\ and\ \citenamefont
		{Kurth}(2013)}]{stefanucci2013kondo}%
	\BibitemOpen
	\bibfield  {author} {\bibinfo {author} {\bibfnamefont {G.}~\bibnamefont
			{Stefanucci}}\ and\ \bibinfo {author} {\bibfnamefont {S.}~\bibnamefont
			{Kurth}},\ }\href@noop {} {\bibfield  {journal} {\bibinfo  {journal} {phys.
				stat. sol. (b)}\ }\textbf {\bibinfo {volume} {250}},\ \bibinfo {pages} {2378}
		(\bibinfo {year} {2013})}\BibitemShut {NoStop}%
	\bibitem [{\citenamefont {Kleeorin}\ and\ \citenamefont
		{Meir}(2017{\natexlab{b}})}]{Kleeorin:PRB:2017}%
	\BibitemOpen
	\bibfield  {author} {\bibinfo {author} {\bibfnamefont {Y.}~\bibnamefont
			{Kleeorin}}\ and\ \bibinfo {author} {\bibfnamefont {Y.}~\bibnamefont
			{Meir}},\ }\href@noop {} {\bibfield  {journal} {\bibinfo  {journal} {Phys.
				Rev. B}\ }\textbf {\bibinfo {volume} {96}},\ \bibinfo {pages} {045118}
		(\bibinfo {year} {2017}{\natexlab{b}})}\BibitemShut {NoStop}%
	\bibitem [{\citenamefont {Stefanucci}\ and\ \citenamefont {van
			Leeuwen}(2013)}]{stefanucci2013nonequilibrium}%
	\BibitemOpen
	\bibfield  {author} {\bibinfo {author} {\bibfnamefont {G.}~\bibnamefont
			{Stefanucci}}\ and\ \bibinfo {author} {\bibfnamefont {R.}~\bibnamefont {van
				Leeuwen}},\ }\href@noop {} {\emph {\bibinfo {title} {Nonequilibrium Many-Body
				Theory of Quantum Systems: A Modern Introduction}}}\ (\bibinfo  {publisher}
	{Cambridge University Press},\ \bibinfo {year} {2013})\BibitemShut {NoStop}%
\end{thebibliography}
\end{document}